\documentclass[12pt,preprint]{aastex}

\shorttitle{[O IV] and X-ray Luminosity Correlation}
\shortauthors{Mel\'endez et. al}

\begin{document}
\title{New Indicators for AGN Power: The Correlation  Between [\ion{O}{4}] $\lambda$25.89$\mu$m and  Hard X-ray Luminosity for 
 Nearby Seyfert Galaxies
}

\author{M. Mel\'endez, S.B. Kraemer and B.K Armentrout}
\affil{Institute for Astrophysics and Computational
Sciences, Department of Physics, The Catholic University of America,
 Washington, DC}
\email{07melendez@cua.edu}
\author{R.P. Deo and D.M. Crenshaw}
\affil{Department of Physics and Astronomy, Georgia State University, Atlanta, GA}
\author{H.R. Schmitt}
\affil{Remote Sensing Division, Naval Research Laboratory, Washington, DC}
\and
\affil{Interferometrics, Inc., Herndon, VA}
\author{R.F. Mushotzky, J. Tueller and  C.B. Markwardt}
\affil{NASA/Goddard Space Flight Center, Greenbelt , MD}
\and
\author{L. Winter}
\affil{University of Maryland, College Park, MD}

\begin{abstract}
We have studied the relationship between the [\ion{O}{4}] $\lambda$25.89$\mu$m emission line luminosities, obtained  from {\it Spitzer} spectra,
  the X-ray continua in the 2-10 keV band, primarily from {\it ASCA}, and the 14-195 keV band  obtained  with the {\it SWIFT}/Burst Alert Telescope (BAT), 
for a sample of nearby (z $<$ 0.08) Seyfert galaxies. For comparison, we have examined the relationship between the  [\ion{O}{3}] $\lambda$5007, the 2-10 keV and the 14-195 keV luminosities for the same set of objects. We find that both the  [\ion{O}{4}] and  [\ion{O}{3}] luminosities are well-correlated with the BAT luminosities. On the other hand, the  [\ion{O}{3}] 
luminosities are better-correlated with 2-10 keV luminosities than are those of  [\ion{O}{4}]. When comparing [\ion{O}{4}] and  [\ion{O}{3}] luminosities for the different 
types of galaxies, we find that the Seyfert 2's have significantly lower [\ion{O}{3}] to [\ion{O}{4}] ratios than the Seyfert  1's. We suggest that this is due to more reddening of the narrow line region (NLR) of the Seyfert  2's. Assuming Galactic dust to gas ratios, the average amount of extra reddening corresponds to a hydrogen column density of $\sim$ few times $10^{21} {\rm cm^{-2}}$, which is a small fraction of the X-ray absorbing columns in the Seyfert 2's. The combined effects of reddening and the X-ray absorption are the probable reason why  the [\ion{O}{3}] versus 2-10 keV correlation is better than the [\ion{O}{4}] versus 2-10 keV, since the  [\ion{O}{4}] $\lambda$25.89$\mu$m emission line is much less affected by extinction. We  present  a grid of photoionization models used to calculate  the physical conditions present in the [\ion{O}{4}] region.  We find that the [\ion{O}{4}] comes from  higher ionization states and lower density regions  than previous studies had determined for  [\ion{O}{3}]. Overall, we find the [\ion{O}{4}] to be an accurate and truly isotropic indicator of the power of the AGN. This suggests that it can be useful in deconvolving the contribution of the AGN and starburst to the spectrum  of Compton-thick and/or X-ray weak sources.

\end{abstract}


\section{Introduction}
The presence or absence of broad optical emission lines has been historically used to separate  Seyfert galaxies into two classes: Seyfert 1 galaxies 
with  broad permitted  and narrow forbidden lines and  Seyfert 2 galaxies with only narrow permitted and forbidden line emission \citep{1974ApJ...192..581K}.
Using spectropolarimetry, \cite{1985ApJ...297..621A} found broad permitted line emission in the Seyfert 2 \objectname{NGC 1068} galaxy, characteristic of a Seyfert 1 spectrum.
These observational results provided the first evidence in favor of a unified model. In this model 
 \citep{1993ARA&A..31..473A} both types of Seyfert galaxies are intrinsically the same with the differences being the 
visibility of the central engine. A geometrically and optically-thick dusty molecular torus-like structure surrounds the central source, as well as the broad line region  (BLR). Therefore the visibility of the nuclear engine  depends on  the viewing angles. 

Following the unified model,  our line of sight to  Seyfert 2 galaxies is obstructed by optically thick material corresponding to 
column densities of $N_H > 10^{22}$ ${\rm cm^{-2}}$ \citep{1999ApJ...522..157R}. For column densities $N_H \leq$ $10^{24}$ ${\rm cm^{-2}}$, photons above few keV can penetrate the torus creating 
an unobstructed view of the nuclear source. One refers to such cases as ``Compton thin". For values of a few times $10^{24}$ ${\rm cm^{-2}}$, only high energy X-ray 
emission (tens of  keV) can pass through the obscuring material \citep{1997ApJ...488..164T}. For  $N_H > 10^{25}$ ${\rm cm^{-2}}$, even high energy  X-rays, above a few tens of keV, are Compton scattered  and the nuclear source 
is completely hidden from our direct view.  Therefore, in order to calculate the intrinsic luminosity for an absorbed object with $N_H > 10^{22}$ ${\rm cm^{-2}}$, 
we need to find an indirect method \citep{1994ApJ...436..586M}. Many authors have used purportedly  isotropic indicators, such as  the  [\ion{O}{3}] $\lambda$5007  line 
(hereafter [\ion{O}{3}]), the infrared continuum, and   the 2-10 keV hard X-ray band \citep{2005ApJ...634..161H,2006A&A...453..525N,1999AJ....118.1169X,1997MNRAS.288..977A,2006A&A...457L..17H}.  \cite{1999ApJS..121..473B} presented a three-dimensional diagram 
for Seyfert 2 galaxies suitable to identify Compton-thick sources, using  $K_\alpha$ iron emission line equivalent width and the  2-10 keV hard X-ray flux normalized to the [\ion{O}{3}] line flux, with the latter corrected for  extinction and  assumed to be a true indicator of the  intrinsic luminosity of the source.

 In this paper, we present and discuss the use of [\ion{O}{4}] $\lambda$25.89$\mu {\rm m}$ (hereafter [\ion{O}{4}]) as an isotropic quantity avoiding  the limitations of previous methods. Since  it has a  relatively high ionization potential
 (54.9 eV), it is less affected by star formation and is significantly less affected by extinction than  [\ion{O}{3}] ($A_{v}\sim 5-45$ corresponds to
 $A_{{\rm 25.89}\mu{\rm m}} \sim 0.1-0.9$) \citep{1998ApJ...498..579G}. Also,  [\ion{O}{4}]  represents an improvement 
over the use of infrared continuum given the difficulty in isolating the AGN continuum from the host galaxy emission \citep{2004A&A...418..465L}. For the X-ray,   we choose the 
{\it Swift} Burst Alert Telescope (BAT)  high Galactic latitude ($|b| > 19^o$) survey in the 14-195 keV band  \citep{2005ApJ...633L..77M}. The survey covers the whole sky 
at $(1 - 3) \times 10^{-11} {\rm ergs}$ ${\rm cm^{-2}}$ $s^{-1}$ and represents a complete  sample including Compton thin AGNs that were missed 
from previous X-ray surveys in the 2-10 keV band because of their high  column densities ($N_H \sim 10^{24}$ ${\rm cm^{-2}}$). Furthermore, Compton-thick 
(i.e.,  $N_H$  $\sim$ few $10^{24}$ ${\rm cm^{-2}}$) sources, which cannot be detected in the 2-10 keV band by the  
{\it Advanced Satellite for Cosmology and Astrophysics} ({\it ASCA}), {\it Chandra} or {\it XMM-Newton},  have been detected by {\it Swift}/BAT
 \citep{2005ApJ...633L..77M,bat}.

 In the present work we compare the $L_{[{\rm O ~ IV}]}$ $L_{14-195 {\rm keV}}$ to $L_{[{\rm O ~ III}]}$ $L_{14-195 {\rm keV}}$ and  $L_{[{\rm O ~ IV}]}$ $L_{2-10 {\rm keV}}$ 
to $L_{[{\rm O ~ III}]}$ $L_{2-10 {\rm keV}}$ relations for a sample of X-ray selected
nearby Seyfert Galaxies. We then use the obtained relations in combination with  photoionization modeling to place constraints on the
physical properties in these emitting regions. We also discuss the different mechanisms behind these correlations and the 
possibilities of using such correlations as a way to find the intrinsic luminosity of AGNs.

\section{Sample and data analysis}

 Using the preliminary results  from the first 3 months of the
 {\it Swift}/BAT high Galactic latitude survey in the 14-195 keV band \citep{2005ApJ...633L..77M} and 
the new results from the 9 month survey \citep{bat} we have compiled a sample of
40 nearby Seyfert galaxies (z $<$ 0.08, except \objectname{3C273} with z=0.16) for which the 2-10 keV, [\ion{O}{4}] $\lambda$25.89$\mu {\rm m}$ and [\ion{O}{3}] $\lambda$5007  fluxes have been measured.

The 2-10 keV fluxes are mainly from {\it ASCA} observations and were
 retrieved from the TARTARUS data base  \citep{2001AIPC..599..991T}, except where noted 
in Tables 1 and 2. All the fluxes have been corrected for Galactic absorption and in the case of multiple observations, the mean flux was taken.
 The  luminosities  in the   [\ion{O}{3}]  line
 were compiled from the literature and are  presented here without  reddening corrections, in Tables~1 and 2.  We present
  [\ion{O}{4}]  fluxes  found in the literature and from  our analysis of unpublished 
archival spectra observed with the Infrared Spectrograph (IRS) \cite[see][]{2004ApJS..154...18H}
 on board the {\it Spitzer Space Telescope} in the ${\rm 1^{st}}$ Long-Low 
(LL1, $\lambda$ = 19.5 - 38.0 $\mu$m, 10.7$"$ $\times$ 168$"$)  IRS order   in  Staring mode. The [\ion{O}{4}] luminosities 
are presented without reddening corrections.

For the analysis of the archival {\it Spitzer} data we used the basic calibrated data (BCD) files preprocessed using the S15.3 IRS pipeline. This includes ramp fitting, dark sky subtraction,
drop correction, linearity correction and wavelength and flux calibrations. The one-dimensional (1D) spectra were extracted from the
IRS data  using the SMART v6.2.4 data processing and analysis package \citep{2004PASP..116..975H}. For the extraction we used 
  the ``Automatic Tapered Column Point Source extraction method" for  LL observations of point sources.
 For  the LL staring mode data we performed  sky subtraction by subtracting
 the BCDs between the two nodes  after we created  median BCDs from each node position. After that, the spectra from each node 
 position for LL1 were averaged to obtain the final spectrum for that order. We performed the line fit with SMART 
using a polynomial to fit the  continuum  and a Gaussian for the line profile.
\objectname{NGC 6300} is the only source, within our sample,  observed in high resolution with the Infrared Spectrograph (IRS)
 in the  Long-High order (LH, $\lambda$ = 18.7 - 37.2 $\mu$m, 11.1$"$ $\times$ 22.3$"$) in Staring mode. For the extraction we used 
  the ``Full" extraction method for  LH observations of point sources. We created median BCDs from each node. Then the spectra from each node 
 position for LH were averaged to obtain the final spectrum. Since background observations are not available  for this galaxy and we required only the 
 [\ion{O}{4}] line flux, we did not perform any background subtraction for this object.


We note the difficulty in  deconvolving  the adjacent [\ion{Fe}{2}] $\lambda$25.99$\mu {\rm m}$ line from the [\ion{O}{4}] in 
low-resolution  IRS spectra. We assumed  that the fluxes measured in low-resolution  mode are mostly from [\ion{O}{4}] given the fact that
the [\ion{Fe}{2}] is due primarily to star formation \citep{2004ApJ...614L..69H,2005ApJ...633..706W,2006ApJ...641..795O}, and  we have a 
 hard X-ray BAT selected sample which is  pre-selected  to be sources  in which the X-ray emission is dominated by an AGN. 
In the next section we will discuss the possible contribution of [\ion{Fe}{2}] to the [\ion{O}{4}] fluxes, based on two  cases within our sample:
the edge-on Seyfert 2 galaxy  \objectname{NGC 3079}  with a nuclear starburst having a low AGN contribution \citep{2007ApJ...655L..73G} 
 and  another Seyfert 2, \objectname{Mrk 3}, with little or no starburst contribution \citep{2007ApJ...671..124D}. 

All the flux errors are within 10\% as obtained from the reduction package {\it SMART} and
from the literature. 

\section{Results}

\subsection{Reddening in the NLR}

\begin{figure}
\epsscale{.80}
\plotone{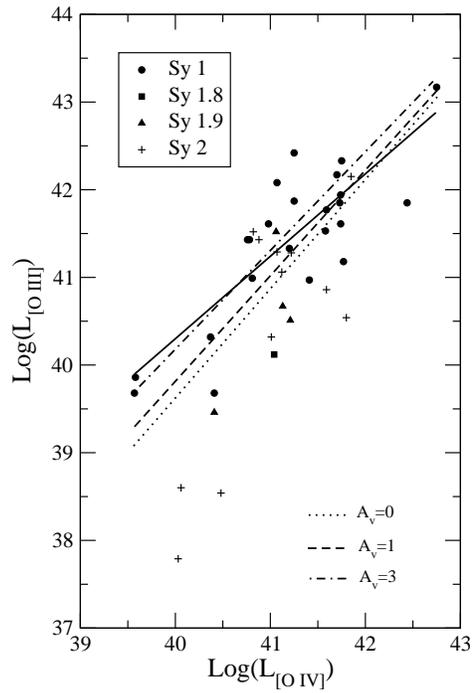}
    \caption{ Comparison between [\ion{O}{4}] and [\ion{O}{3}] luminosities. The solid line represents the linear regression calculated for Seyfert 1 galaxies; the 
 dotted, dashed and point-dashed lines represents the linear regression for the  sample with  the Seyfert 2 galaxies  corrected by extinction in the visible with 
$A_v=0,1$ and $3$, respectively. For comparison purposes we group Seyfert 1.8's and 1.9's with the  Seyfert 2's.
\label{fig1}}
\end{figure}

In Figure ~\ref{fig1} we compare the [\ion{O}{3}] and the [\ion{O}{4}]  luminosities.
For our purposes, we group  Seyfert  1.2's and 1.5's  with  Seyfert 1's.
From the plot we notice that Seyfert  2, Seyfert 1.9 and Seyfert 1.8 galaxies show lower [\ion{O}{3}] luminosities than Seyfert 1 galaxies. The mean 
value of the luminosities  our sample are:
for Seyfert 1's  $\log (L_{[{\rm O~ III}]})=41.4 \pm 0.9$ and $\log (L_{[{\rm O~ IV}]})=41.2 \pm 0.8$, for  Seyfert 1.9's $\log (L_{[{\rm O~ III}]})=40.5 \pm 0.8$ and 
$\log (L_{[{\rm O~ IV}]})=40.9 \pm 0.4$; and for Seyfert 2's $\log (L_{[{\rm O~ III}]})=40.0 \pm 1.0$ and $\log (L_{[{\rm O~ IV}]})= 41.0  \pm 0.6$.
A linear regression  
for each individual group yields a relation of the type: for Seyfert 1's $L_{[{\rm O~ III}]} \propto L_{[{\rm O~ IV}]}^{0.9 \pm 0.1}$;
Seyfert 1.9's $L_{[{\rm O~ III}]} \propto L_{[{\rm O~ IV}]}^{2 \pm 1}$ and for Seyfert 2's $L_{[{\rm O~ III}]} \propto L_{[{\rm O~ IV}]}^{1.8 \pm 0.5}$. These results are consistent with previous reports  of additional  reddening  in the narrow line region (NLR) of Seyfert 2's \cite[e.g.,][]{1991MNRAS.250..422J,1994ApJ...436..586M,1994A&A...283..791K,2005ApJ...620..151R,2006A&A...453..525N}.

 Using the extinction law derived by \cite{1989ApJ...345..245C}, the 
estimated extinction in the visible is  $A_{v}\sim 1-6$ mag, calculated from 
the mean values for the [\ion{O}{3}]  luminosity  required  to ``correct" the Seyfert 2's (Sy1.8/1.9/2's)
[\ion{O}{3}] to those from Seyfert 1's, assuming a ratio of total to
 selective extinction of $R_{v}=3.1$ (which represents a typical value for the diffuse interstellar medium). 
In Figure~\ref{fig1} we present  these findings by comparing  the linear regression obtained for  Seyfert 1 galaxies (solid line), with those obtained for all the sample with the [\ion{O}{3}] luminosities corrected for extinction in Seyfert 2 galaxies. 
Therefore, we interpreted  the  higher $\log (L_{[{\rm O~IV}]}/L_{[{\rm O~III}]} \ga 1$ found in Seyfert 2 galaxies to be a result of  reddening.

There is a relative absence of  Seyfert 1's in nearly edge-on host galaxies \citep[see][]{1980AJ.....85..198K,1995ApJ...454...95M,2001ApJ...555..663S}. Our sample
is consistent with this trend, in that  we find that Seyfert 2's  have, on average, more inclined host galaxies with a ratio of b/a$=0.5\pm0.3$\footnote{The values for the major and minor diameter of the host galaxy, a and b respectively, were taken from NED} contrasting with b/a$=0.7\pm0.2$ for  Seyfert 1 galaxies. These results are also supported by the observed correlation between continuum reddening and inclination of the host galaxy \citep{2001ApJ...562L..29C}. Thus, 
host galaxy-related obscuration may contribute to the inclination dependence of the [\ion{O}{3}] emission, although our sample is too small to draw strong conclusions.


Assuming a gas-to-dust ratio for the host galaxy of  $5.2 \times 10^{21}\rm{cm^{-2}mag^{-1}}$ \citep{1985ApJ...294..599S} the additional absorbing column calculated  from 
the  values derived for  extinction ($A_v$) is in the range of $N_H \sim (2 - 10) \times 10^{21} {\rm cm^{-2}}$ for the Seyfert 2 galaxies. The median  values
 for the X-ray  hydrogen column densities for our Seyfert 2 and Seyfert 1 samples are $N_H = 2.1 \times 10^{23} \rm{cm^{-2}}$  and $N_H = 1.2  \times 10^{21} \rm{cm^{-2}}$,
 respectively \citep{bat}. The observed discrepancy between the X-ray column density and those required for the extinction of the [\ion{O}{3}] emission lines is consistent  with 
previous results   \citep[e.g.,][]{1982ApJ...257...47M,1985ApJ...296...69R,2001A&A...365...28M}. The X-ray column density
 is measured along the line of sight to the nucleus and the derived reddening measures  the column density towards the NLR. Thus while different gas-to-dust ratios
 may contribute,  it is also likely that there is an  additional attenuating  gas component closer to the X-ray source which is not affecting the [\ion{O}{3}] emission, as first suggested by \cite{1982ApJ...257...47M}.

\subsection{The Correlation Between [\ion{O}{3}],[\ion{O}{4}] and the 2-10 keV Band}
\begin{figure*}
\epsscale{.80}
\plotone{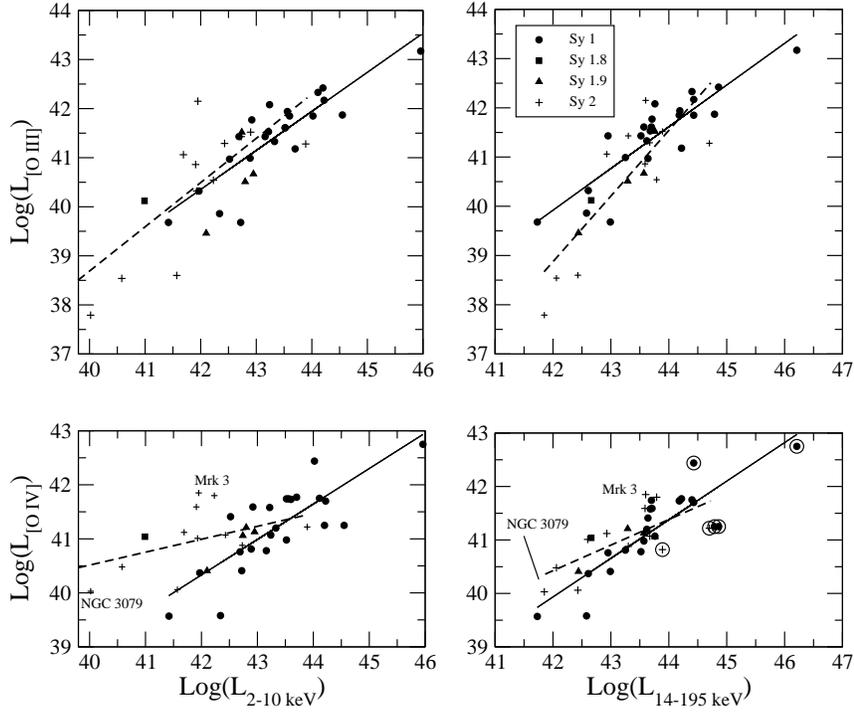}
    \caption{Correlation between [\ion{O}{4}]  and [\ion{O}{3}] luminosities with hard X-ray  (2-10 keV) band and BAT 
 (14-195 keV) luminosities. The solid line represents the linear regression calculated  for the Seyfert 1 galaxies and the dashed line represents the linear regression for  the Seyfert 2 galaxies. The circles in the lower right panel  represent radio loud (RL) objects.
\label{fig2}}
\end{figure*}

In Figure ~\ref{fig2}  we present  comparisons  between [\ion{O}{4}] and [\ion{O}{3}] luminosities and X-ray  (2-10 keV)  and BAT 
(14-195 keV) luminosities. Linear fits to these relations, and statistical analysis  are presented in Table ~\ref{table0}. 
From the different relations, the lowest level of significance for a correlation  is obtained for the [\ion{O}{4}] 
versus the 2-10 keV hard X-ray. We propose that comparing the heavily-absorbed hard X-ray band in Seyfert 2's with the  
``unabsorbed" [\ion{O}{4}]  may have weakened any correlation. In order to corroborate this statement we 
 used the relation,  $L_{[{\rm O~ IV}]} \propto L_{2-10 {\rm keV}}^{0.7\pm 0.1}$,  derived from the Seyfert 1's  to estimate the hard X-ray 2-10 keV fluxes for 
the Seyfert 2 galaxies using  their observed [\ion{O}{4}]. To investigate the effect of ionized absorption on the 2-10 keV X-ray continuum, we created a multiplicative table (mtable) model using the ``grid" and ``punch XSPEC" options \citep{2006PASP..118..920P} in the photoionization code CLOUDY, version 07.02.01, last described by \cite{1998PASP..110..761F}.  A single zone  and  simple power law incident continuum model was assumed, with photon index $\Gamma = {\rm 1.8}$, and low- and high-energy cutoffs at 1 micron and 100 keV, respectively.  Hydrogen column density was then varied, with hydrogen number density fixed at $n_H = 10^8~{\rm cm^{-3}}$, and ionization parameter $\log (U) = -1$, where the ionization parameter  $U$ is defined as \cite[see][]{2006agna.book.....O}:
\begin{equation}
U=\frac{1}{4\pi R^2cn_H}\int^\infty_{\nu_o}\frac{L_\nu}{h\nu}d\nu=\frac{Q(H)}{4\pi R^2cn_H},
\label{u}
\end{equation}
where R is the  distance to the cloud 
, c is the speed of light  and $Q(H)$ is the flux of ionizing photons.

The resulting table was then fed into the X-ray spectral fitting package XSPEC (version 12.3.1).  We reproduced the incident power law in XSPEC using the POWERLAW additive model, attenuated by absorption from the Cloudy-produced mtable.  Using the XSPEC ``flux" command, we determined the effect of various hydrogen column densities on the emergent 2-10 keV X-ray flux. Finally, we obtained the column densities needed in order to correct the observed hard X-ray 
fluxes with their intrinsic counterpart derived from the [\ion{O}{4}].

\begin{figure}
\epsscale{.80}
\plotone{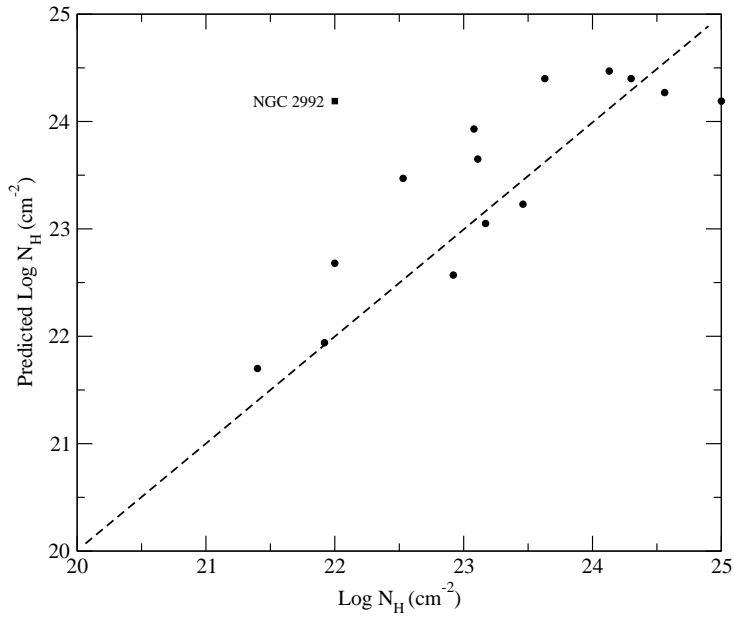}
    \caption{ Comparison  between the predicted absorbing column densities and values 
found in the literature for our Seyfert 2 population. The predicted values were derived using  the [\ion{O}{4}] and 2-10 keV luminosity correlation found 
for Seyfert 1's. The square represents the column density prediction based on 
ASCA observations for the  
 X-ray flux  in 2004 for NGC~2992 (see text for details). The dashed line represents a 1:1 correspondence.\label{fig3}}
\end{figure}

In Figure ~\ref{fig3} we compare the predicted absorbing column densities with 
values found in the literature (Table ~\ref{table_column}). Even though we used  a simple X-ray model, the predicted  column densities are in good agreement
 with those from the literature. \objectname{NGC 2992} is the evident outlier from our sample. The X-ray flux has been seen to vary dramatically throughout the history of \objectname{NGC 2992} (Trippe et al., in preparation). As we mentioned before, the 2-10 keV fluxes are mainly from {\it ASCA} observations (in order to maintain consistency within our sample). {\it ASCA} observations 
 for this object were taken in 1994 and {\it Spitzer} IRS staring observations in 2005. Using the hard X-ray flux,
 $f_{2-10keV} = 8.88 \times 10^{-11}{\rm ergs~cm^{-2}~s^{-1}}$, observed by \cite{2007ApJ...666...96M} with the {\it Rossi X-ray Timing Explorer} (RXTE) in 2005,
  we obtained better agreement with the absorbing column density 
 predicted from the  [\ion{O}{4}] (see Figure~\ref{fig3} and Table~\ref{table_column}). Overall, these results confirm the heavy absorption present in Seyfert 2 galaxies and the effectiveness of [\ion{O}{4}] as a true indicator of the intrinsic X-ray luminosity of the galaxy.


%
%
The good empirical correlation between the [\ion{O}{3}] emission line and the hard X-ray in the 2-10 keV band has been discussed extensively
\citep[e.g.,][]{1997MNRAS.288..977A,1999AJ....118.1169X,2005ApJ...634..161H,2006A&A...453..525N}. This correlation is often used to determine the 
intrinsic X-ray luminosity, both in absorbed  AGN where the observed X-ray flux is affected  by absorption. We corroborate this in our sample, in that
we find a strong linear correlation between [\ion{O}{3}] and  2-10 keV luminosities. However, caution must be taken when using this correlation since 
 [\ion{O}{3}] and hard X-ray luminosities are essentially absorbed quantities, especially  in Seyfert 2 sources. We suggest that  the
  combined effect of  extinction in the  [\ion{O}{3}]  luminosities ($A_{v}\sim 1-6$ mag) with correspondingly  heavily absorbed  2-10 keV fluxes
 for Seyfert 2 galaxies   results in a false correlation, thereby  underestimating the  true  intrinsic luminosity  of the AGN, particularly
 for nearly X-ray Compton thick sources.

\subsection{The Correlation Between [\ion{O}{3}], [\ion{O}{4}] and the BAT Band}

Based on the  statistical analysis we found  equally good  correlations for the  [\ion{O}{3}] and [\ion{O}{4}] relations with the 14-195 keV band. As we mentioned before, the [\ion{O}{4}] represents a relatively   uncontaminated 
 quantity and therefore  the relation with an unabsorbed quantity such as the BAT fluxes should give the best correlation, assuming 
that the [\ion{O}{4}] is a good indicator of the power of the AGN. However,  from
the statistical  analysis, this does not seem to be the case.
 To explore this behavior we applied  linear regression fits to the individual Seyfert 1 and Seyfert 2 galaxies in our sample (see Figure~\ref{fig2}).  

The highest level of correlation was obtained for the Seyfert 1 galaxies:  [\ion{O}{3}] versus the 14-195 keV  ($r_s=0.88$, $P_r=1.2 \times 10^{-7}$) and  
 [\ion{O}{4}] versus the 14-195 keV ($r_s=0.84$,  $P_r=3.6 \times 10^{-7}$). As we found previously, [\ion{O}{3}] in Seyfert 2 galaxies is  
more affected by extinction than  in Seyfert 1's. This could  create  a well-correlated linear distribution when comparing with the 14-195 keV BAT fluxes only if there is similar  attenuation present in the 14-195 keV band for Seyfert 2's. Following the individual statistical analyses,  Seyfert 2 objects have less correlated  [\ion{O}{3}] 
versus  14-195 keV hard X-ray  ($r_s=0.73$, $P_r=1.6 \times 10^{-2}$) and  [\ion{O}{4}] versus  14-195 keV hard X-ray luminosities
  ($r_s=0.64$, $P_r=3.4 \times 10^{-2}$) than  Seyfert 1 galaxies. Partial absorption of the  hard X-ray fluxes, especially in  Seyfert 2 galaxies, with the relatively  
unobscured [\ion{O}{4}] is suggested as the cause for the low level  of correlation. We found that, using  $L_{[{\rm O~ III}]} \propto L_{14-195 {\rm keV}}^{0.9 \pm 0.1}$ derived from the Seyfert 1 population, we underpredicted the BAT luminosities for
the Seyfert 2 galaxies. This is an expected result considering  the extinction of  the  [\ion{O}{3}] emission. On the other hand,
 using the $L_{[{\rm O~ IV}]} \propto L_{14-195 {\rm keV}}^{0.7\pm 0.1}$ derived from the Seyfert 1 population, we   overpredicted the BAT luminosities for most of the
 Seyfert 2 galaxies. There are two different scenarios to explain the over-prediction: an overestimation of [\ion{O}{4}] due to contamination from
[\ion{Fe}{2}] (see Section 2)  and  absorption in the 14-195 keV BAT band.

Although we cannot dismiss the former, we expect minimal starburst 
contribution in our X-ray selected sample. However, we are aware of the importance of [\ion{Fe}{2}] emission especially in weak  X-ray sources.
In order to estimate the  [\ion{Fe}{2}] contamination coming from the starburst contribution we calculated  the predicted  [\ion{O}{4}] from the relation with the 14-195 keV luminosities 
for  Seyfert 1 galaxies and compared with the observed  [\ion{O}{4}] luminosities for two extreme Seyfert 2 objects (see Figure ~\ref{fig2}): \objectname{NGC 3079}, with a major contribution from the starburst \citep{2007ApJ...655L..73G},
and \objectname{Mrk 3}, with no starburst \citep{2007ApJ...671..124D}.  Using these luminosities we estimated an upper limit for the [\ion{Fe}{2}]  starburst contribution
to be  $\sim 1.5$ times the uncontaminated [\ion{O}{4}] luminosity for  the low-resolution spectra of \objectname{NGC 3079}. 
Comparing the full low-resolution with the full high-resolution IRS spectra obtained by \cite{2005ApJ...633..706W} we clearly see  [\ion{Fe}{2}] as the dominant 
component  (and the impossibility of resolving the [\ion{O}{4}] in the low-resolution spectra for \objectname{NGC 3079}). We also noticed that the strong [\ion{O}{4}] emission line in \objectname{Mrk 3} shows  no contribution from [\ion{Fe}{2}]. However, the nearly X-ray Compton-thick \objectname{Mrk 3} \citep{2005MNRAS.360..380B}, with  uncontaminated and unabsorbed [\ion{O}{4}], still overpredicts the measured 
BAT flux and the observed [\ion{O}{4}] of \objectname{NGC 3079} is consistent with the X-ray flux, if the source is nearly X-ray Compton-thick \citep{2006A&A...446..459C}, as we showed with the XSPEC simulation.

 Therefore, the most likely scenario to explain the hard X-ray overprediction in Seyfert 2's  is  partial   absorption in the 14-195 keV BAT band. In order to investigate this last statement further we used, as before,  the X-ray spectral fitting package XSPEC. For energies above 10 keV, we employed the standard XSPEC model PLCABS, which  simulates attenuation of a power law 
continuum by dense, cold matter \citep{1997ApJ...479..184Y}.  The model grid has a high energy flux calculation limit of 50 keV, so our flux range was limited to 14-50 keV.  To accommodate large column densities, we set the maximum number of scatterings considered by the model to a value of 12, as suggested by \cite{1997ApJ...479..184Y}.  As before, the power law photon index was set to $\Gamma={\rm 1.8}$.  Other variable model parameters were set to default values, and the column density was varied to determine relative changes in the  predicted 14-50 keV  X-ray  luminosities using the [\ion{O}{4}]- BAT correlation for Seyfert 1's, and the observed BAT 14-50 keV  X-ray  luminosities of the Seyfert 2 galaxies. From these results, we derived a mean column density of $N_H \approx (3.2 \pm 0.8) \times 10^{24} {\rm cm^{-2}}$ for our Seyfert 2 population.
 Given the  energy limitation of the previous method, this result is in agreement with a derived value of 
$N_H \approx (2.4 \pm 0.8) \times 10^{24} {\rm cm^{-2}}$, calculated using the average  values for the predicted  and the observed  BAT X-ray luminosities, and assuming a purely Thomson cross-section. From  this value of column density  the BAT 14-195 keV X-ray band appear to have been partially absorbed. This result is in good agreement with the large column densities  observed for  \objectname{Mrk 3} and \objectname{NGC 3079} (see Table~\ref{table_column}).

\subsection{Radio Loudness and [\ion{O}{4}]}
There is a well known  linear correlation between the radio and the [\ion{O}{3}] luminosities, and that this correlation is similar for both radio quiet and 
radio loud AGNs \citep[e.g.,][]{1999AJ....118.1169X,2001ApJ...555..650H}. We noticed that some of the  objects in 
our sample, which  show more more scattered with respect to the linear correlation between [\ion{O}{4}] and BAT luminosities,
are radio loud sources (see Figure~\ref{fig2}, open circles). In order to investigate whether or not this is an actual trend or  a sample selection effect,  we expanded our original sample to include more radio loud objects. We used the sample and radio loudness classification from \cite{1999AJ....118.1169X} and selected 
the objects  that have been observed with {\it Spitzer} and are  classified 
 as  Seyfert galaxies (following NED and SIMBAD classification).
These new objects do not have 14-195 keV fluxes from the 9 month BAT survey and therefore are not included in any of the previous calculations 
regarding the correlations of the [\ion{O}{4}] and the hard X-ray. 
 We obtained the [\ion{O}{4}] fluxes from the literature  and {\it Spitzer} IRS staring mode archival (unpublished) spectra in low resolution. The  extended sample is presented in Table~2 (the labels are the same as in Table~1).

\begin{figure}
\epsscale{.80}
\plotone{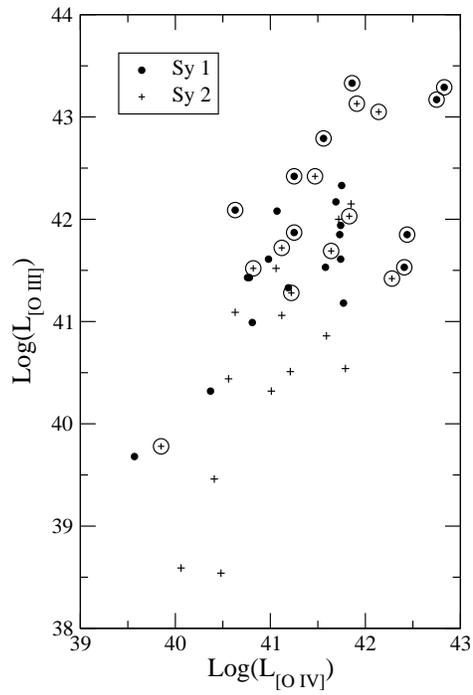}
    \caption{Correlation between [\ion{O}{4}] and [\ion{O}{3}]   luminosities  for the combined sample, i.e. the original and extended sample. 
The circles represent radio loud (RL) objects. For the purpose of comparison we group  Seyfert's 1.8 and 1.9 with the  Seyfert 2 galaxies. \label{fig4}}
\end{figure}

 The relationship  between the [\ion{O}{3}] and [\ion{O}{4}] has been studied before in powerful Fanaroff-Riley class II radio galaxies \citep{2005A&A...442L..39H}. 
\cite{2005A&A...442L..39H} found [\ion{O}{3}] not to be a true indicator of the AGN, because the  optical emission from the NLR suffers extinction. In Figure~\ref{fig4} we compare the [\ion{O}{4}] and [\ion{O}{3}]   luminosities  for the combined sample, i.e. the original and extended sample.  From the comparison  
it become clear  that radio loud sources exhibit  higher emission line luminosities than those of other types of active galaxies. This result is in agreement 
 with  previous findings and could be explained by a proposed bowshock model  where part of the NLR emission  is 
being powered by  a radio-emitting jets,  present in  radio loud AGNs \citep[eg.][]{1992MNRAS.255..351T,1995ApJ...455..468D,1997ApJ...485..112B,1998ApJ...495..680B}.
 Most of the radio loud objects included in the extended sample were chosen to be powerful radio sources \citep{2005A&A...442L..39H,2006ApJ...647..161O}.
 Therefore, our results are hampered by  selection effects. Furthermore,  our small  sample does not allow us to extract any  
statistically significant results to determine  a clear  trend that may distinguish between Type 1 and Type 2 radio loud AGNs. For example, there may be evidence 
for a high [\ion{O}{4}]/[\ion{O}{3}] in radio loud Seyfert 1's, as suggested by \cite{2005A&A...442L..39H}, however the same trend is not 
seen in the RL Seyfert  2 galaxies. This remains as an open question for further analysis.

\section{Photoionization modeling}

To investigate the physical conditions in the emission line regions for both the [\ion{O}{4}] and [\ion{O}{3}], we generated
 single-zone, constant-density models. 
We used a set of roughly  solar abundances \cite[e.g.,][]{1989AIPC..183....1G}. The logs of the abundances relative to H by number are: He: -1; C: -3.46; N: -3.92; O: -3.19;  Ne: -3.96;  Na: -5.69;  Mg: -4.48; Al: -5.53;
Si: -4.50; P: -6.43;  S: -4.82; Ar: -5.40;  Ca: -5.64;  Fe: -4.40 and  Ni: -5.75. We assumed a column density of $10^{21}{\rm cm^{-2}}$, which is  typical
 of the narrow line region \citep[e.g.,][]{2000ApJ...531..278K}. Although it is likely that there is dust mixed in with the emission-line gas in the 
NLRs of Seyfert galaxies \citep[e.g.,][]{1986ApJ...307..478K,1993ApJ...404L..51N}, more recent studies using optical and UV long-slit spectra obtained with the Space Telescope
 Imaging Spectrograph aboard the {\it Hubble Space Telescope} indicate that the dust/gas ratios are significantly lower than in the Galactic ISM 
\citep{2000ApJ...544..763K,2000ApJ...532..256K,2000ApJ...531..278K}. Therefore, we have not included dust in our models.

For the Spectral Energy Distribution (SED)  we assume a broken power law as used  by \cite{2004ApJ...607..794K} and similar 
to that suggested for \objectname{NGC 5548} \citep{1998ApJ...499..719K} and \objectname{NGC 4151} \citep{2000ApJ...531..278K}
  of the form $F_\nu \propto \nu^{-\alpha}$, with $\alpha {\rm =0.5}$ below 13.6 eV, $\alpha {\rm =1.5}$ from 13.6 eV to 1 keV and 0.8 at higher energies.
We generated a grid of photoionization models varying   the ionization parameter $U$ and total hydrogen number density ($n_H$). The average  BAT luminosity  (14-195 keV) of the  sample, 2.67$\times 10^{43}$ ergs ${\rm s^{-1}}$,  
yields a flux of ionizing photons of $Q (H) \sim$1.2$\times 10^{54}$ photons ${\rm s^{-1}}$. 

 In order to study the physical conditions present in the emission-line  regions, we need to
 use intrinsic quantities (free of viewing angle effects). As discussed in the previous section, the [\ion{O}{3}] emission line seems to be  more affected by
 reddening in Seyfert 2 than Seyfert 1 galaxies. Consequently, we  used  the values derived from the  Seyfert 1 galaxies in
 our sample to constrain the photoionization parameters. We suggested that, given the isotropic properties for [\ion{O}{4}], the physical 
conditions derived from our Seyfert 1 population could be extended to Seyfert 2 as well.

 From the Seyfert 1's we find that the average  [\ion{O}{4}] and [\ion{O}{3}] ratio  is $\sim 1.0 \pm 0.2$. 
Assuming that the [\ion{O}{3}] and  [\ion{O}{4}] emission comes from the same gas, we used the generated  grid of models to find the range in $U$ and $n_H$ for 
which we can obtain a ratio of approximately  unity for different values of  extinction ($A_v$). We obtained
  a range for the ionization parameter  $-1.50 < \log (U) < -1.30$ and for the  
 hydrogen density,  $ 2.0 < \log (n_H) < 4.25~{\rm (cm^{-3})}$, assuming no extinction ($A_v=0$). The
  values for $n_H$ lie around the [\ion{O}{4}]  critical density, $\log(n_c)\sim 3.7$, where the line intensity peaks.  The critical density for the  $^2{\rm P}^o_2$ level
 of ${\rm O^{3+}}$, which is the upper level of the [\ion{O}{4}] line, was calculated using the formalism described in \cite{2006agna.book.....O} using  atomic data from \cite{1992ApJS...80..425B} and \cite{1998A&AS..131..499G}.

\cite{2005MNRAS.358.1043B} found from their single-zone approximation a relatively small range for the values of $n_H$ 
 ($\log(n_e)\sim 5.85 \pm 0.7 ~{\rm cm^{-3}}$)\footnote{ The photoionization code CLOUDY uses the total hydrogen density ($n_H$) instead of the 
electron density ($n_e$) as used by \cite{2005MNRAS.358.1043B}. We assumed, in our simple model, both densities to be roughly the same.}
 and a relatively small range of $U$  ($\log (U)= -3 \pm 0.5$) where both the [\ion{O}{3}] and H$\beta$ are emitted efficiently.
For their two-zone model assuming  $n_e = 10^{3} {\rm cm^{-3}}$ for the outer zone and $n_e = 10^{7} {\rm cm^{-3}}$ and $\log (U) = -1$ for the inner zone, they found  
higher  $U$ values for the outer zone than in the single-zone approximation  ($-3.5 < \log (U) < -2$). These
 are typical values for the conditions in the NLR  \citep[e.g.,][]{2000ApJ...531..278K}. While our models predicted higher ionization parameters than 
\cite{2005MNRAS.358.1043B}, they did not attempt to include [\ion{O}{4}].

\begin{figure}
\epsscale{.80}
\plotone{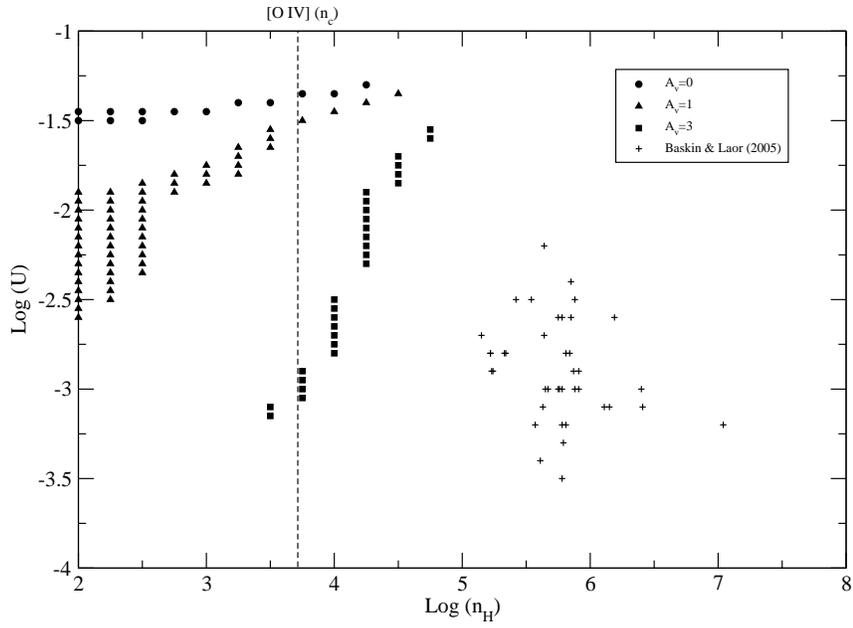}
    \caption{ Comparison between the parameter space required to obtain a ratio of approximately  unity for [\ion{O}{4}]/[\ion{O}{3}] for different extinction magnitudes for the [\ion{O}{3}].
 For comparison we shown the parameter space obtained by \cite{2005MNRAS.358.1043B}.\label{fig5}}
\end{figure}

In Figure \ref{fig5}, we present  the required parameter space to obtain a [\ion{O}{4}]/[\ion{O}{3}] of unity for different values of $A_v$ and compared 
 with that found by \cite{2005MNRAS.358.1043B} in their single-zone model. This comparison shows the 
dependence of this ratio in terms of the ionization parameter, hydrogen density and extinction. For a higher extinction, the parameter space required is closer
 to that found by \cite{2005MNRAS.358.1043B}, meaning that the intrinsic ratio is smaller than unity and therefore representing  a region where the [\ion{O}{3}] is 
dominant, relative to the [\ion{O}{4}] emission. In the absence of such large extinction ($A_v \ga 6 {\rm mag}$), it is likely that
the [\ion{O}{4}] is coming from a more highly  ionized and lower density gas. Our comparison of [\ion{O}{4}] and [\ion{O}{3}] show that  the single-zone model cannot  reproduce the physical conditions present in the NLR gas since $U$  would be too low to produce the observed [\ion{O}{4}] and the line would be collisionally suppressed.
On the other hand, our results are in better agreement with the higher ionization state found in their constant density  outer zone. However, this
higher ionization and lower density cloud population cannot fully reproduce the observed [\ion{O}{4}].

Since $U$ depends on the flux of ionizing photons ($Q(H)$), the distance to the source ($R$) and the 
hydrogen density of the gas ($n_H$), we used  the model grid predictions to  calculate a range for the  distances to the source. Then, we determined a range for 
the  covering factors ($Cf$) by adjusting to the observed [\ion{O}{4}] luminosity of the sample. We found a distance of the emitting gas from the central source of 
$250~{\rm pc} > R >  30~{\rm pc}$ and a range of covering factors  of $ 0.04 < Cf < 0.15$. The range in parameter space in our grid was dominated by matter bounded models. In the matter bounded case, all the material is ionized and some fraction of the ionizing radiation passes through the gas unabsorbed. Therefore, the covering factor must be higher than that from a radiation bounded model (in which all the ionizing radiation is absorbed in the NLR), to account for the observed strength of the emission lines. However, the inclusion of dust in the models will decrease the size of the ionized zone, which would further increase the predicted covering factors \citep[e.g.,][]{1993ApJ...404L..51N,2005MNRAS.358.1043B}. If the [\ion{O}{4}] arose in such dusty gas, it would imply higher covering factors than our models predict. 
Nevertheless, the results from our dust-free models are in good agreement with the $Cf$ ($\sim 0.02 -0.2$) obtained  by \cite{2005MNRAS.358.1043B} in their 
 radiation bounded dusty models. There is  evidence for a component of high ionization matter  bounded gas
in the NLR in which optical lines such as [\ion{Ne}{5}]$\lambda$3426 and [\ion{Fe}{7}] $\lambda$6087 form \citep[e.g.,][]{1996A&A...312..365B,2000ApJ...532..256K,2000ApJ...531..278K}, so it is likely that the [\ion{O}{4}] originates in the same gas, while most of the [\ion{O}{3}]
 arises in conditions similar to those found by \cite{2005MNRAS.358.1043B}.


\begin{figure}
\epsscale{.80}
\plotone{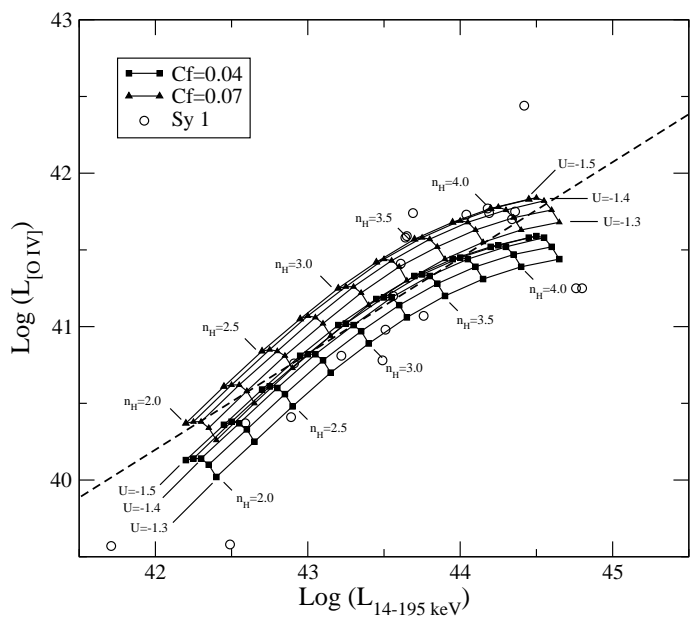}
    \caption{ Comparison  between the correlation of [\ion{O}{4}]  luminosities with BAT X-ray luminosities
 and a grid of photoionization models for different covering factors (Cf) for a distance of $R=130~{\rm pc}$. The dashed line represents the linear regression obtained for the observed luminosities for  all the  sample and the open circles represent the observed Seyfert 1's luminosities (see Figure~\ref{fig2}).
$n_H$ refers to the $\log$ of the hydrogen nucleus density and $U$ to the $\log$ of the ionization parameter. The vertical dashed line indicates the critical 
density for the [\ion{O}{4}] line \label{fig6}}
\end{figure}

In Figure~\ref{fig6} we present the photoionization  model  grid overlaid on  the Seyfert 1 luminosities  
  and the linear regression for all the sample.  Each case corresponds to a different set of covering factors ($Cf$) for a distance from the ionizing source to the gas cloud of $R=130~{\rm pc}$. The model X-ray luminosities are calculated using U, the fixed radial distance, the range in densities shown in Figure~\ref{fig6}, and our assumed 
SED (see Equation~1).

As presented in Figure ~\ref{fig6}, our  simple photoionization model can reproduce the observed $L_{[{\rm O~IV]}}$ to $L_{14-195 {\rm keV}}$ relationship.
 In order to fit the  higher luminosity sources, the models predict  that 
 more distant or  higher density gas is required for the NLR. We dismiss the latter because, at high densities (i.e. higher than the critical density for 
[\ion{O}{4}]), the [\ion{O}{4}] is  collisionally de-excited  and the line intensity decreases. Therefore, the most likely scenario is a more distant 
NLR for high luminosity sources. On the other hand, Figure ~\ref{fig6} shows that  our simple photoionization model underpredicts  [\ion{O}{4}] at $R=130~{\rm pc}$, even at low densities.  Consequently, in the low luminosity sources the NLR must be fairly small ($R < 150$ pc) with a bigger covering factor ($Cf \ga 0.07$), since gas at or below the critical density of the [\ion{O}{4}] would not be sufficiently ionized at large distances from the source. These results suggests that  the size of the NLR scales with the luminosity of the source, in agreement with the observed good correlation between the NLR size and the luminosity of the AGN \citep{2003ApJ...597..768S}.


\section{Conclusions}
We have explored the relationship between the $L_{[{\rm O~ IV}]}$-$L_{14-195~keV}$ and related correlations for a sample of X-ray selected
nearby Seyfert Galaxies. For the X-ray selected sample the  [\ion{O}{4}]  and the [\ion{O}{3}]  luminosities are
 well-correlated with the hard X-ray luminosity in the 14-195 keV band over a range of $\sim$ 3 and 5 orders of magnitude, respectively.

From the [\ion{O}{4}] and [\ion{O}{3}] comparison we derived an  absorbing column density from the reddening toward the NLR for Seyfert 2 galaxies, 
$N_H \sim (2 - 10) \times 10^{21} {\rm cm^{-2}}$, which is smaller than the median value for the X-ray column densities, $N_H = 2.1 \times 10^{23} \rm{cm^{-2}}$. The differences in the line of sight for the different absorbing columns are consistent with an additional attenuating gas component
close to the X-ray source. 
We also suggested that part of the obscuration in the NLR of Seyfert 2 galaxies may occur in the disk of the host galaxy, given the fact that, in our sample, 
Seyfert 2 galaxies are on average more inclined.

We obtained a  correlation between the $L_{[{\rm O~ IV}]}$-$L_{2-10~keV}$ that is not as strong as the one between $L_{[{\rm O~ III}]}$-$L_{2-10~keV}$,
 which results from comparing  the heavily absorbed  2-10 keV band, 
especially in Seyfert 2 galaxies, with the  intrinsic  (unabsorbed) [\ion{O}{4}] fluxes. As pointed out in previous works, there is a 
good correlation between [\ion{O}{3}]  and hard X-ray 2-10 keV luminosities. However, we  conclude that  the
combined effects of reddening in the [\ion{O}{3}]  and absorption in the  2-10 keV band may  result in a false correlation.

For the $L_{[{\rm O~ III}]}$ and $L_{[{\rm O~ IV}]}$ relations with the $L_{14-195{\rm keV} }$ BAT
 band we found equally good correlations. Using the linear regression found for Seyfert 1 galaxies  we 
underpredict the hard X-ray 14-195 keV BAT luminosities of  Seyfert 2's   obtained from  the $L_{[{\rm O~ III}]}$ versus hard 
X-ray  $L_{14-195 {\rm keV} }$ relation and overpredict   the hard X-ray 14-195 keV BAT luminosities using 
the $L_{[{\rm O~ IV}]}$ relation. We explain the former case due to the extinction of  the [\ion{O}{3}] ($A_{v}\sim 1-6$ mag) and the latter with 
absorption in the  BAT band. We found that for a column density of $ N_H \sim 3  \times 10^{24} \rm{cm^{-2}}$ the  BAT luminosities  
could  be smaller, by a factor of $\sim 5 \pm 3$,  than the  intrinsic luminosities derived from the Seyfert 1's relation. 

Assuming that the [\ion{O}{3}] and  [\ion{O}{4}] come from the same gas,  and using the mean fluxes of the sample, we conclude that the emitting region for [\ion{O}{4}]
 is  in a higher ionization state and lower density  than  the components modeled for the [\ion{O}{3}] emitting region by \cite{2005MNRAS.358.1043B} and has a
 mean covering factor  of $Cf \sim 0.07$. The physical conditions 
derived from the photoionization models indicate that [\ion{O}{4}]  originates in the inner $\sim$ 150 pc of the NLR.

In conclusion, we propose  the [\ion{O}{4}] as  a truly isotropic property of AGNs given its high ionization potential and that it is 
 unaffected by reddening,   which makes this line easy to identify and extract. This is true at least in galaxies with
 minimal or no star formation as for our X-ray selected sample. Where  star formation activity  has an 
important contribution this may not be the case, especially with contamination from [\ion{Fe}{2}] $\lambda$25.99$\mu {\rm m}$ in low-resolution IRS spectra.
In a future work we will expand our sample to include the complete 22 month {\it SWIFT}/BAT survey and high resolution {\it Spitzer} IRS observations to
continue to explore these open issues.

\acknowledgments
We would like to thank our anonymous referee for her/his insightful comments and suggestions  that improved the  paper. 
This research has made use of the NASA/IPAC Extragalactic Database (NED) which is operated by the Jet Propulsion Laboratory, 
California Institute of Technology, under contract with the National Aeronautics and Space Administration. The IRS was a collaborative venture between 
Cornell University and Ball Aerospace Corporation funded by NASA through the Jet Propulsion Laboratory and Ames Research Center.
SMART was developed by the IRS Team at Cornell University and is available through the Spitzer Science Center at Caltech. Basic research in astronomy at the NRL is supported by 6.1 base funding.

\clearpage

\bibliographystyle{apj}

\bibliography{ms}
\clearpage

\begin{deluxetable}{llllllll}
\rotate
\tabletypesize{\scriptsize}
\tablewidth{0pt}
\tablecaption{Sample of Seyfert Galaxies}
\tablehead{ && \multicolumn{5}{c}{Luminosities (ergs ${\rm s^{-1}}$)}\\ \colhead{Name} & \colhead{Type} &\colhead{$z$}& \colhead{$\log(L_{[{\rm O~ III}]})$} & \colhead{$\log(L_{[{\rm O~ IV}]})$} &
 \colhead{$\log(L_{2-10 {\rm keV}})$} & \colhead{$\log(L_{14-195 {\rm keV}})$ } & \colhead{Remarks}}
\startdata

3C120	&	1	&	0.033010	&	41.85\tablenotemark{a}	&	42.44	&	44.02	&	44.43	&	L\\
3C273	&	1	&	0.158339	&	43.17\tablenotemark{a}	&	42.75	&	45.96	&	46.21	&	L\\
3C382	&	1	&	0.057870	&	41.87\tablenotemark{b}	&	41.25	&	44.55	&	44.79	&	L\\
3C390.3	&	1	&	0.056100	&	42.42\tablenotemark{b}	&	41.25	&	44.20	&	44.86	&	L\\
3C452	&	2	&	0.081100	&	41.28\tablenotemark{b}	&	41.22	&	43.89	&	44.70	&	L\\
3C84	&	2	&	0.017559	&	41.52\tablenotemark{c}	&	40.82	&	42.90	&	43.89	&	L\\
Circinus	&	2	&	0.001448	&	38.54\tablenotemark{c}	&	40.48\tablenotemark{l}	&	40.58	&	42.06	&Q	\\
ESO103-G035	&	1.9	&	0.013286	&	40.67\tablenotemark{d}	&	41.13	&	42.95	&	43.57	&\nodata	\\
IC4329a	&	1.2	&	0.016054	&	41.18\tablenotemark{d}	&	41.77	&	43.70	&	44.22	&Q	\\
Mrk   6	&	1.5	&	0.018813	&	41.77\tablenotemark{e}	&	41.59\tablenotemark{k}	&	42.92	&	43.71	&\nodata	\\
Mrk  79	&	1.2	&	0.022189	&	41.61\tablenotemark{e}	&	41.74\tablenotemark{k}	&	43.52	&	43.7	&Q	\\
Mrk 348	&	2	&	0.015034	&	41.29\tablenotemark{e}	&	41.07\tablenotemark{k}	&	42.43	&	43.67	&\nodata	\\
Mrk 3	&	2	&	0.013509	&	42.15\tablenotemark{e}	&	41.85	&	41.95	&	43.6	&Q	\\
Mrk 509	&	1.2	&	0.034397	&	42.33\tablenotemark{f}	&	41.75	&	44.11	&	44.40	&Q	\\
MCG-6-30-15	&	1.5	&	0.007749	&	39.68\tablenotemark{m}	&	40.41\tablenotemark{k}	&	42.72	&	42.99	&\nodata	\\
MCG-2-58-22	&	1.5	&	0.046860	&	42.17\tablenotemark{h}	&	41.70	&	44.22	&	44.43	&Q	\\
NGC 1365	&	1.8	&	0.005457	&	40.12\tablenotemark{d}	&	41.04	&	40.99	&	42.66	&\nodata	\\
NGC 2992	&	2	&	0.007710	&	41.06\tablenotemark{i}	&	41.12	&	41.69	&	42.93	&Q	\\
NGC 3079	&	2	&	0.003723	&	37.79\tablenotemark{j}	&	40.03\tablenotemark{k}	&	40.02	&	41.85	&I	\\
NGC 3227	&	1.5	&	0.003859	&	40.32\tablenotemark{e}	&	40.37\tablenotemark{k}	&	41.97	&	42.61	&Q	\\
NGC 3516	&	1.5	&	0.008836	&	40.99\tablenotemark{e}	&	40.81\tablenotemark{k}	&	42.89	&	43.25	&Q	\\
NGC 3783	&	1.5	&	0.009730	&	41.43\tablenotemark{g}	&	40.78	&	43.16	&	43.52	&Q	\\
NGC 4051	&	1.5	&	0.002336	&	39.68\tablenotemark{g}	&	39.57\tablenotemark{k}	&	41.42	&	41.73	&Q	\\
NGC 4151	&	1.5	&	0.003319	&	41.43\tablenotemark{g}	&	40.76\tablenotemark{k}	&	42.69	&	42.95	&Q	\\
NGC 4388	&	2	&	0.008419	&	40.86\tablenotemark{g}	&	41.59	&	41.91	&	43.59	&Q	\\
NGC 4507	&	1.9	&	0.011801	&	41.52\tablenotemark{g}	&	41.06	&	42.74	&	43.76	&Q	\\
NGC 526A	&	1.5	&	0.019097	&	41.33\tablenotemark{g}	&	41.20\tablenotemark{k}	&	43.33	&	43.62	&Q	\\
NGC 5506	&	1.9	&	0.006181	&	40.51\tablenotemark{g}	&	41.21	&	42.80	&	43.29	&Q	\\
NGC 5548	&	1.5	&	0.017175	&	41.61\tablenotemark{e}	&	40.98\tablenotemark{k}	&	43.52	&	43.57	&Q	\\
NGC 6240	&	2	&	0.024480	&	40.54\tablenotemark{c}	&	41.80	&	42.23	&	43.79	&Q	\\
NGC 6300	&	2	&	0.003699	&	38.60\tablenotemark{c}	&	40.06	&	41.57	&	42.43	&Q	\\
NGC 7172	&	2	&	0.008683	&	41.43\tablenotemark{d}	&	40.88\tablenotemark{k}	&	42.73	&	43.30	&\nodata	\\
NGC 7213	&	1.5	&	0.005839	&	39.86\tablenotemark{d}	&	39.58	&	42.34	&	42.58	&\nodata	\\
NGC 7314	&	1.9	&	0.004763	&	39.46\tablenotemark{g}	&	40.41\tablenotemark{k}	&	42.10	&	42.44	&Q	\\
NGC 7469	&	1.5	&	0.016317	&	41.53\tablenotemark{g}	&	41.58	&	43.22	&	43.68	&Q	\\
NGC 7582	&	2	&	0.005254	&	40.32\tablenotemark{g}	&	41.01	&	41.93	&	42.60	&Q	\\
NGC 931	&	1.5	&	0.016652	&	40.97\tablenotemark{d}	&	41.41\tablenotemark{k}	&	42.52	&	43.64	&\nodata	\\
NGC 985	&	1	&	0.043143	&	41.94\tablenotemark{g}	&	41.74	&	43.56	&	44.19	&Q	\\
PG1501+106	&	1.5	&	0.036420	&	41.85\tablenotemark{h}	&	41.73	&	43.60	&	44.18	&Q	\\
PG1534+580	&	1	&	0.029577	&	42.08\tablenotemark{g}	&	41.07	&	43.24	&	43.76	&Q	\\

\label{table1}
\enddata
\tablecomments{The luminosities were calculated from the fluxes using  $H_o=71{\rm km}$${\rm s^{-1}}$${\rm Mpc^{-1}}$ and  a deceleration parameter $q_o=0$ with  redshift values
 taken from NED. The [O~IV] luminosities obtained in the present work  are presented in the table without references.
 The 2-10 keV fluxes are from ASCA observations and were retrieve from the TARTARUS data base \citep{2001AIPC..599..991T} except for:
\objectname{3C 84} from {\it XMM-Newton} observations \citep{2006ApJ...642...96E};
 \objectname{3C 452} from {\it Chandra} ACIS observations \citep{2002ApJ...580L.111I};\objectname{MCG -2-58-22} and  \objectname{NGC 6300} from 
{\it BeppoSAX}  MECS observations \citep{2007A&A...461.1209D};\objectname{Mrk 79} from {\it XMM-Newton} \citep{2006MNRAS.365..688G}; 
\objectname{Mrk 79}, which is from an {\it Ariel 5} observations \citep{1978MNRAS.183..129E}; \objectname{NGC 3079} which is from {\it XMM-Newton} \citep{2006A&A...446..459C}. The 14-195 keV fluxes presented in this table are from the 9 months {\it SWIFT}/BAT high Galactic latitude survey \citep{bat} except for:
\objectname{Circinus} and \objectname{NGC 3079}, which are from Tueller et al. (in preparation). The last column follow the radio loudness
 classification from \cite{1999AJ....118.1169X}, L= Radio loud, Q= Radio quiet and I= Intermediate.}  
\tablerefs{(a) \cite{1993MNRAS.263..999T}, (b) \cite{1989MNRAS.240..701R}, (c) \cite{1996ApJS..106..399P} , (d) \cite{1997ApJ...486..132B} and references therein
, (e) \cite{1975ApJ...199...19A}, (f) \cite{1980ApJ...241..894Y}, (g) \cite{1992ApJS...79...49W} and references therein,(h) \cite{1988ApJS...67..249D}, (i) \cite{1980MNRAS.193..563W},
(j) \cite{1997ApJS..110..321B},(k) \cite{2007ApJ...671..124D},(l) \cite{2002A&A...393..821S}, (m)\cite{1992MNRAS.257..677W}.
}
\end{deluxetable}

\clearpage

\begin{deluxetable}{lllllll}
\tabletypesize{\scriptsize}
\tablewidth{0pt} 
\tablecaption{Extended Sample}
\tablehead{ && \multicolumn{4}{c}{Luminosities (ergs ${\rm s^{-1}}$)}\\ \colhead{Name} & \colhead{Type} & \colhead{$z$}&\colhead{$\log(L_{[{\rm O~ III}]})$} & 
\colhead{$\log(L_{[{\rm O~ IV}]})$} &\colhead{$\log(L_{2-10 {\rm keV}})$}&\colhead{Remarks}}
\startdata 

3C 109	&	 1.8	&0.305600&	43.13\tablenotemark{a}	&	41.91\tablenotemark{j}	&	45.19	&L	\\
3C 192	&	 2	&0.059709&	41.72\tablenotemark{a}	&	41.12	&	\nodata	&L	\\
3C 234	&	 1	&0.184800&	43.29\tablenotemark{a}	&	42.83\tablenotemark{k}	&	44.19	&L	\\
3C 249.1&	 1	&0.311500&	43.33\tablenotemark{b}	&	41.86\tablenotemark{j}	&	44.79	&L	\\
3C 321	&	 2	&0.096100&	41.69\tablenotemark{c}	&	41.64\tablenotemark{j}	&	42.78	&L	\\
3C 323.1&	 1	&0.264300&	42.80\tablenotemark{d}	&	41.56\tablenotemark{j}	&	45.01	&L	\\
3C 33	&	 2  	&0.059700&	42.03\tablenotemark{a}	&	41.83\tablenotemark{k}	&	43.79	&L	\\
3C 381	&	 1	&0.160500&	41.53\tablenotemark{a}	&	42.41\tablenotemark{k}	&	\nodata&L	\\
3C 433	&	 2	&0.101600&	41.42\tablenotemark{e}	&	42.28\tablenotemark{k}	&	44.08	&L	\\
3C 445	&	 1	&0.056200&	42.09\tablenotemark{f}	&	40.63\tablenotemark{j}	&	43.76	&L	\\
3C 459	&	 2	&0.219900&	42.42\tablenotemark{f}	&	41.47\tablenotemark{j}	&	\nodata	&L	\\
3C 79	&	 2	&0.255900&	43.05\tablenotemark{a}	&	42.14\tablenotemark{j}	&	\nodata	&L	\\
Mrk 231	&	 1	&0.042170&	41.94\tablenotemark{g}	&	41.42\tablenotemark{l}	&	43.76	&I	\\
Mrk 573	&	 2	&0.017179&	42.00\tablenotemark{h}&	     41.72\tablenotemark{m}	&	41.25	&Q	\\
NGC 5643&	 2	&0.003999&	40.44\tablenotemark{h}	&	40.56\tablenotemark{m}	&	40.35	&Q	\\
NGC 6251&	 2	&0.024710&	39.78\tablenotemark{i}	&	39.85	&	42.28	&L	\\
TOL 0109-383&	 2	&0.011764&	41.09\tablenotemark{i}	&	40.63\tablenotemark{m}	&	41.72	&Q	\\

\label{table2}
\enddata
\tablecomments{The luminosities were calculated from the fluxes using  $H_o=71{\rm kms^{-1}Mpc^{-1}}$ and  a deceleration parameter $q_o=0$ with  redshift values
 taken from NED. The [O~IV] luminosities obtained in the present work  are presented in the table without references. The 2-10 keV fluxes are from ASCA observations and were retrieve from the TARTARUS data base \citep{2001AIPC..599..991T} except for:
\objectname{3C 33} from {\it XMM-Newton} observations (\cite{2006ApJ...642...96E}); \objectname{Mrk 573} from   {\it XMM-Newton}  observations 
\citep{2005A&A...444..119G} and \objectname{TOL 0109-383} from \cite{2005ApJ...634..161H}. The last column follow the radio loudness
 classification from \cite{1999AJ....118.1169X}, L= Radio loud, Q= Radio quiet and I= Intermediate.}
\tablerefs{(a) \cite{1989MNRAS.240..701R}, (b) \cite{1991MNRAS.250..422J}, (c) \cite{1991ApJS...75.1011G}, (d) \cite{1996ApJS..102....1B}, (e) \cite{1981ApJ...250..469S},
(f) \cite{1993MNRAS.263..999T}, (g) \cite{1988ApJS...67..249D}, (h) \cite{1992ApJS...79...49W}, (i) \cite{1996ApJS..106..399P}, (j) \cite{2005A&A...442L..39H},
 (k) \cite{2006ApJ...647..161O},(l) \cite{2007ApJ...656..148A},(m) \cite{2002A&A...393..821S}.
}

\end{deluxetable}

\clearpage

\begin{deluxetable}{lcccc}
\tabletypesize{\small}
\tablewidth{0pt} 
\tablecaption{Linear regressions and statistical analysis for the [\ion{O}{4}] and [\ion{O}{3}] to their X-ray luminosities}
\tablehead{  &\multicolumn{2}{c}{$\log ({\rm L_{2-10 keV}})$} & \multicolumn{2}{c}{$\log ({\rm L_{14-195 keV}})$}\\ 
             &\colhead{a} &\colhead{b}&\colhead{a}&\colhead{b}}
\startdata 
$\log {\rm L_{[O~ III]}}$&0.82626&5.6382&1.0771&$-$5.8703\\
                        &$0.8\pm 0.1$&$5 \pm 4$&$1.1 \pm 0.1$&$-6 \pm 5$\\
&\multicolumn{2}{c}{$r_s=0.791$, $P_r=7.8 \times 10^{-7}$}&\multicolumn{2}{c}{$r_s=0.825$, $P_r=2.5 \times 10^{-7}$}\\
$\log {\rm L_{[O~ IV]}}$&0.38903&24.467&0.61487&14.358\\
                        &$0.4 \pm 0.1$&$24 \pm 3$&$0.6\pm 0.1$&$14 \pm 3$\\
&\multicolumn{2}{c}{$r_s=0.596$, $P_r=2.0 \times 10^{-4}$}&\multicolumn{2}{c}{$r_s=0.798$, $P_r=6.3 \times 10^{-7}$}\\

\enddata
\label{table0}
\tablecomments{a and b represent the regression coefficient (slope) and regression constant (intercept) respectively. $r_s$ is the
Spearman rank order correlation coefficient and $P_r$ is the null probability. For each relation we presented the exact linear regression values,
 the values as constrained by their statistical errors and the Spearman rank and null probability.}
\end{deluxetable}

\clearpage

\begin{deluxetable}{lcc}
\tabletypesize{\small}
\tablewidth{0pt} 
\tablecaption{Column densities for Seyfert 2 galaxies.}
\tablehead{ \colhead{Name} &\colhead{$\log N_H$ (XSPEC)}&\colhead{$\log N_H$}}
\startdata 

Circinus& 24.27 &24.56\tablenotemark{a}\\
ESO103-G035 & 23.05 & 23.17\tablenotemark{b}\\
Mrk 348 & 23.65 & 23.11\tablenotemark{c}\\
Mrk 3 & 24.47 & 24.13 \tablenotemark{d}\\
NGC 2992 & 22.68& 22.00 \tablenotemark{e}\\
NGC 3079 & 24.19 & 25.00 \tablenotemark{f}\\
NGC 4388 & 24.40 & 23.63 \tablenotemark{g}\\
NGC 4507 & 23.23 & 23.46 \tablenotemark{g}\\
NGC 5506 & 23.47 & 22.53 \tablenotemark{g}\\
NGC 6240 & 24.40 & 24.30 \tablenotemark{h}\\
NGC 6300 & 21.70 & 21.40 \tablenotemark{i}\\
NGC 7172 & 22.57 & 22.92 \tablenotemark{b}\\
NGC 7314 & 21.94 & 21.92 \tablenotemark{b}\\
NGC 7582 & 23.93 & 23.08 \tablenotemark{g}\\
\enddata
\label{table_column}
\tablecomments{The XSPEC calculations are based on the comparison between [O~IV] predictions and observed hard X-ray 2-10 keV fluxes. See text for details.}
\tablerefs{ (a) \cite{1999A&A...341L..39M}, (b) ASCA observations retrieved using TARTARUS data base, (c) \cite{2002MNRAS.332L..23A}, (d) \cite{2005MNRAS.360..380B},
 (e) \cite{2000A&A...355..485G}, (f) \cite{2006A&A...446..459C}, (g) \cite{1999ApJS..121..473B}, (h)\cite{1999A&A...349L..57V}, (i)\cite{2007A&A...461.1209D}
}
\end{deluxetable}

\end{document}